\documentclass[12pt,preprint]{aastex}
\begin{document}
   \title{The identification of physical close galaxy pairs}

\author{D.S.L. Soares}
\affil{Departamento de F\'\i sica, ICEx, Universidade Federal de Minas Gerais, 
              C.P. 702, 30123-970 Belo Horizonte, MG Brazil \\
              \email{dsoares@fisica.ufmg.br}
             }

\received{}
\accepted{}

\begin{abstract}

A classification scheme for close pairs of galaxies is proposed. The scheme is 
motivated by the fact that the majority of apparent close pairs are in 
fact wide pairs in three-dimensional space. This is demonstrated by means of 
numerical simulations of random samples of binary galaxies and the scrutiny of 
the resulting projected and spatial separation distributions.

Observational strategies for classifying close pairs according to the scheme 
are suggested. As a result, physical (i.e., bound and spatially) close 
pairs are identified. 

\end{abstract}


%

\section{Introduction}
The investigation of binary galaxies is jeopardized by the intrinsic 
lack of temporal observational tracking of the orbital parameter space  
caused by orbital periods being of the order of hundreds of megayears. 
The alternative is the statistical study of samples of pairs relying on 
reasonable frequency  distributions of the relevant orbital parameters. 
This has been done often since the pioneering work in the field by 
Holmberg \cite{erik}. 
Besides the unknowns mentioned above there is a second major setback: 
binary galaxy catalogues are always contaminated by the so-called 
optical pairs, i.e., unphysical, unbound pairs seen in 
projection on the plane of the sky. Earlier works, including Holmberg's, dealt
with the problem by imposing restrictive selection criteria in the
catalogues (e.g., Page et al. 1961; Karachentsev 1972, 1987; Turner 1976). 
Consequently the projected linear separations of pairs in
these catalogues are typically 50 kpc. If one intends to study the
distribution of galaxy mass to larger spatial extents such samples are
clearly not adequate. A large sample of wide pairs is important in the
determination of the size of dark halos, believed to exist within and around the
visible parts of galaxies. Later attempts have been made in this direction
(van Moorsel 1982; Schweizer 1987; Soares 1989; Charlton \& Salpeter 1991; 
Chengalur et al. 1993; Nordgren et al. 1998) with the addition of wide pairs 
with separations as large as 1 Mpc, and even more. The contamination by optical 
pairs remained a fundamental issue in all of these works.

Here I focus on a definite group of pairs, namely, close pairs, to investigate 
the important point, often neglected,  that \emph{closeness in the plane of 
the sky is not always a guarantee of three-dimensional closeness}. 

\begin{figure}
   \centering
\includegraphics[width=\textwidth]{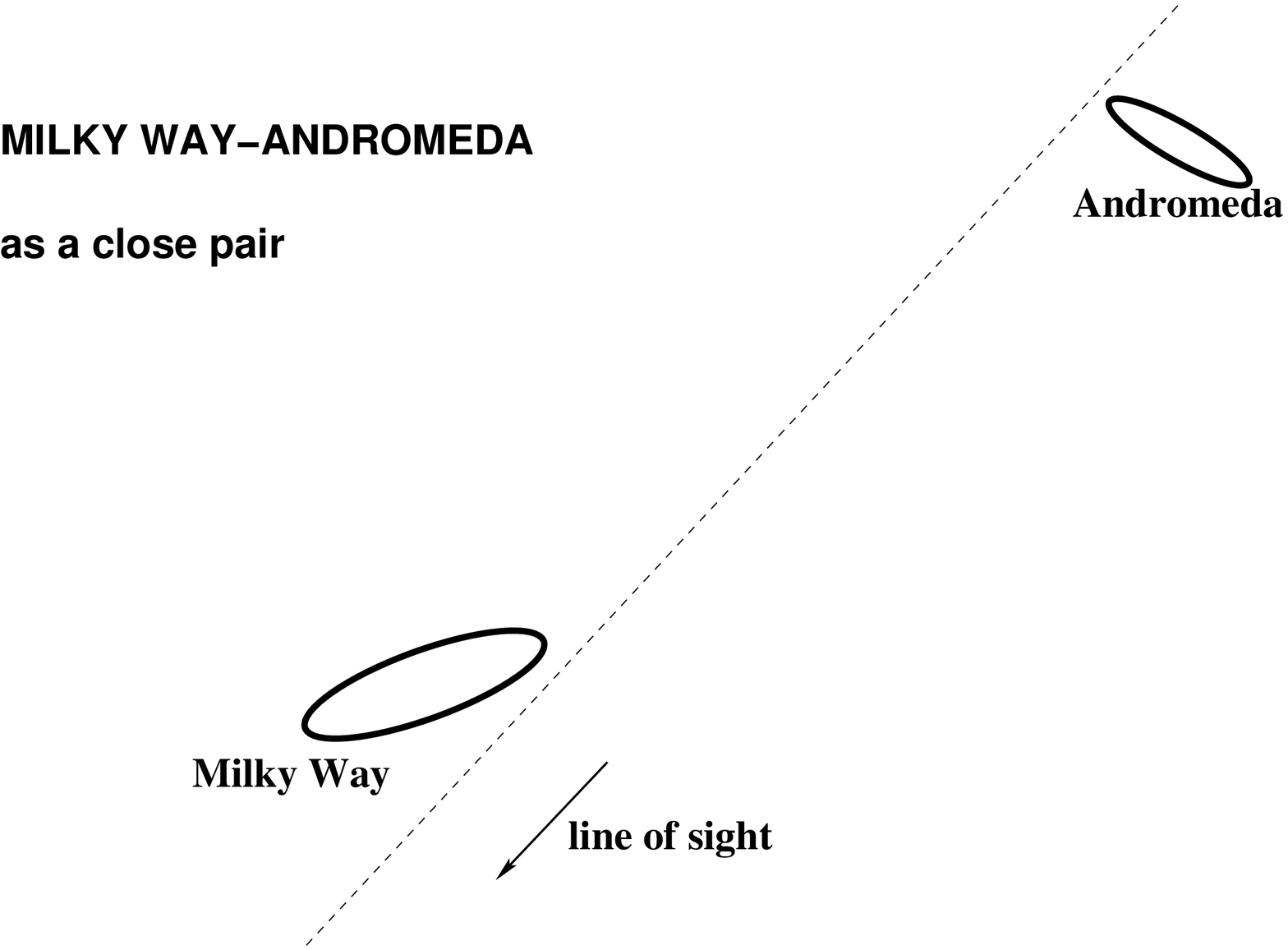}
      \caption{MW-And system and a line of sight 
towards a hypothetical distant observer. 
              }
         \label{mwand}
   \end{figure}

An example from our backyard is useful as an illustration. Let us 
consider the most trivial galaxy pair: the Milky Way--Andromeda 
system (MW-And). Being the dominant galaxies in the Local Group, 
these galaxies form a wide pair with 700 kpc of separation.
The cartoon in Figure \ref{mwand} shows the MW-And pair and a line of sight 
pointing to the outskirts of the Virgo Cluster, located at 
about 15 Mpc. An observer at that location --- assuming the Andromeda galaxy
has the same linear dimension as the Milky Way ---
would notice a difference of less than 5\% in their angular size. He could well
classify the pair as a close one. With a velocity difference of 119 km s$^{-1}$ 
at the line of sight considered, the MW-And pair would be a strong
candidate to be in a list of close pairs. The reality is of another nature:
MW-And is a wide pair. Its apogalacticon separation might be estimated 
at about 2 Mpc, if the system is on a highly-eccentric orbit and the galaxies 
have approximately 10$^{12}$ $M_\odot$. Most plausible, the system has recently 
reached its apocentric configuration in a slightly-eccentric orbit, implying 
a significant transverse velocity at present. The proper motion 
of M31 has indeed been constrained to about 100 km s$^{-1}$ by Loeb et al. 
\cite{loeb}.

\clearpage

%
%

\clearpage

Recent papers aimed at investigating a number of properties of
close pairs, such as star formation (Barton et al. 2000; Woods et al. 2006) 
and merger rates, do not give due relevance to the fact that projected physical 
separation ($R_p$) cannot suffice as a unique constraint for closeness. Patton et al.
(2000, 2002) define a close pair as having 5h$^{-1}$ $< R_p
\leq 20$h$^{-1}$ kpc and an upper limit in the pair radial-velocity difference 
of 500 km s$^{-1}$. The velocity criterion represents a somewhat loose general
constraint on physical association.
They are followed by others in essentially the same procedure
(e.g., Nikolic et al. 2004). I argue here that this is 
just the first step in the selection procedure. It has to be 
complemented by an investigation of the tidal effects on the individual 
galaxies in the pair in order to have a clear definition of closeness.

Other studies address the presence of Markarian activity in galaxy pairs. 
Byrd \& Valtonen \cite{byrd} find that Markarian activity is an indication 
of tidal interaction. Their results should be confronted with the 
closeness of the sample pairs identified by an activity-blind procedure, 
as suggested here.

In section 2 I show, using Monte Carlo simulations, that a small projected 
separation is not a secure and fair indication of a small spatial separation. 
This is a consequence of the fact that two-body orbits have most of their 
phase-space time at apocentric configurations (cf. Kepler's law). The scene 
is thus set for the presentation of the classification scheme  
in section 3. The relevant observational data for sorting 
apparently close pairs throughout classes are suggested in section 4. 

Section 5 summarizes the main conclusions.


\section{Spatial and projected separations in close pairs}
The most simple way of disclosing the three-dimensionality of projected 
separations in binary galaxies is by means of Monte Carlo--type simulations 
of bound pairs. These are generally an useful tool when one wants to analyze an 
observed binary sample in order to extract information such as galaxy mass and 
angular momentum and orbital period and eccentricity. Such a statistical 
method has been used intensively in the past (e.g., Turner 1976; 
van Moorsel 1982; Schweizer 1987; Oosterloo 1988;  
Soares 1990, 1996).

To simulate a sample of binary galaxies a number of input orbital parameters
have to be specified, namely, the distribution of spatial separations and of
orbital eccentricities. Furthermore, a mass model for galaxies has to be
defined, as do projection rules to transform orbital velocity and radial
separation into observed quantities such as line-of-sight velocity difference
and apparent projected separation. 

\subsection{Simulation procedure}
The simulation procedure is established according to the objective one has in 
mind. Here the main goal is to investigate how close pairs, defined by 
their apparent separations, are distributed in three dimensions or, in other 
words, are distributed in the apparent versus spatial separation plane. 
The simulations are case problems that span a wide range of
possibilities in the whole phase-space of binary orbits. 
To reach the goal I generated samples of 2000 artificial pairs with 
the following specifications. 

\begin{description}
\item{(i)} Mass model: two equal-mass galaxies in Keplerian orbits. Galaxies
move under their mutual point-mass Newtonian gravitational potential. I 
further add the restriction that galaxies never get closer than the sum
of their (approximate) half-mass radii. I adopt a conservative value of
10 kpc for the sum. In practice this means that galaxies getting closer
than this limiting separation rapidly merge and are therefore excluded from 
the simulated sample. Self-consistent {\it N}-body simulations of binary-galaxy
orbit secular evolution have conclusively shown that such systems indeed
merge on a time-scale much smaller than a Hubble time (e.g., Bartlett \&
Charlton 1995; Chan \& Junqueira 2001). Evidently, galaxies are 
not point masses and may possess well-extended (dark) mass distributions. 
As long as the binary system is isolated from a third-body gravitational 
influence, and furthermore the mutual gravitational interaction is only 
radially dependent, the point-mass assumption is a reasonable description of 
the general consequences of orbital angular momentum conservation, which 
relies on Kepler's areal law, i.e., on the fact that 
galaxies spend most of their time at apocenter.

\item{(ii)} Spatial separation: samples with either a fixed apocentric
separation $R_{apo}$ of either 200 or 600 kpc. These two values span fairly 
the average (apocentric) separations of binary galaxies as suggested 
by several works (Soares 1989; Schweizer 1987; Charlton 
\& Salpeter 1991; Chengalur 1994; Nordgren 1997). 
A more rigorous simulation should consider a distribution of binary spatial
separations. Earlier work (Gott \& Turner 1979, but see Soares 
1989, p. 17) suggested that binary galaxies follow, down to small
scales, an extension of the classical galaxy two-point correlation function
(Peebles 1980), but more recent evidence points to a multi-variable
correlation function, depending on galaxy type, luminosity and local galaxy
number density. An $R_{apo}$ value smaller than 200 kpc is not considered 
here because, as mentioned above, such pairs merge quickly.

\item{(iii)} Eccentricity distribution: three distributions, spanning the
entire orbital phase-space. They are $f(e)=\delta(0.9), f(e)=2e,$ and
$f(e)=\delta(0)$. If binary galaxies were formed by early capture, suggested
already by Holmberg (1937; see also Schweizer 1987), the triangular
distribution $f(e)=2e$ would be suitable since it allows for some amount of
transverse orbital motion. The extreme cases of highly eccentric orbits
($e=0.9$) and circular orbits bracket the triangular distribution and are
considered for completeness and reveal fruitful consequences (see
below).

\item{(iv)} Projection on the sky: the normal to the orbital plane is
distributed at random \emph{in space}, that is, the orbital inclination $i$
has a distribution $f(i)\propto \sin(i)$.
\end{description}
%
\begin{table}
\begin{minipage}[t]{\columnwidth}
\caption{Number of simulated close pairs  out of 2000} 
\label{nclose}
\vspace{.5cm}
\centering
\renewcommand{\footnoterule}{}  
\begin{tabular}{c c c c }        
\hline\hline                 
$R_{apo}$& $f(e)=\delta(0.9)$ & $f(e)=2e$ & $f(e)=\delta(0)$ \\    
         &Median $R_p$ & Median $R_p$ & Median $R_p$ \\
\hline
200 kpc & 284  & 228  & 51  \\      
        & 121 kpc    & 125 kpc    & 173 kpc    \\
\hline
600 kpc & 67  & 30  & 4  \\
        & 360 kpc    & 360 kpc    & 520 kpc    \\
\hline
\end{tabular}
\end{minipage}
\end{table}

A pair orbit is simulated with prescriptions 1--4; velocities
and radial separations are drawn at a random time instant within the orbit,
and, from prescription 4, the projected separation onto the plane of the sky 
and the corresponding line-of-sight pair velocity difference are calculated.
The random choice of orbital phase implies, from Kepler's
law, that pairs at apocenter --- large separations --- are naturally
favored. For each pair of $R_{apo}$ and $f(e)$ the process is repeated until a
list of 2000 pairs is simulated, and therefrom I investigate the
correlation between projected and spatial separation.
%
%
\begin{table}
\begin{minipage}[t]{\columnwidth}
\caption{Percentage of close pairs with $R>50$ kpc}       
\label{rplus}      
\vspace{.5cm}
\centering
\renewcommand{\footnoterule}{}  
\begin{tabular}{c c c c}        
\hline\hline                 
$R_{apo}$& $f(e)=\delta(0.9)$ & $f(e)=2e$ & $f(e)=\delta(0)$ \\    
\hline                        
200 kpc & 52.5\% & 69.3\% & 100\% \\      
600 kpc & 53.7\% & 73.3\% & 100\% \\
\hline                                   
\end{tabular}
\end{minipage}
\end{table}
%
\begin{table}
\caption{Classification of close pairs }
\label{clas}      
\vspace{.5cm}
\centering                          
\begin{tabular}{l c c }        
\hline\hline                 
Class & Pair  & Interaction \\    
\hline                        
CP I   & merger & strongest \\      
CP II  & tide-loud & strong \\
CP III & tide-quiet & weak\\
CP IV  & optical  & weakest \\
\hline                                   
\end{tabular}
\end{table}
%

\subsection{Projected versus spatial separation}
A \emph{putative} close pair is defined as a pair of 
galaxies having at most the fiducial projected separation $R_{p,max}=$ 50 
kpc, which is typical in the studies mentioned above. I then ask: \emph{How 
many pairs with $R_p\le R_{p,max}$ have spatial separations $R > R_{p,max}$?}

\begin{figure}
   \centering
\includegraphics[width=12cm]{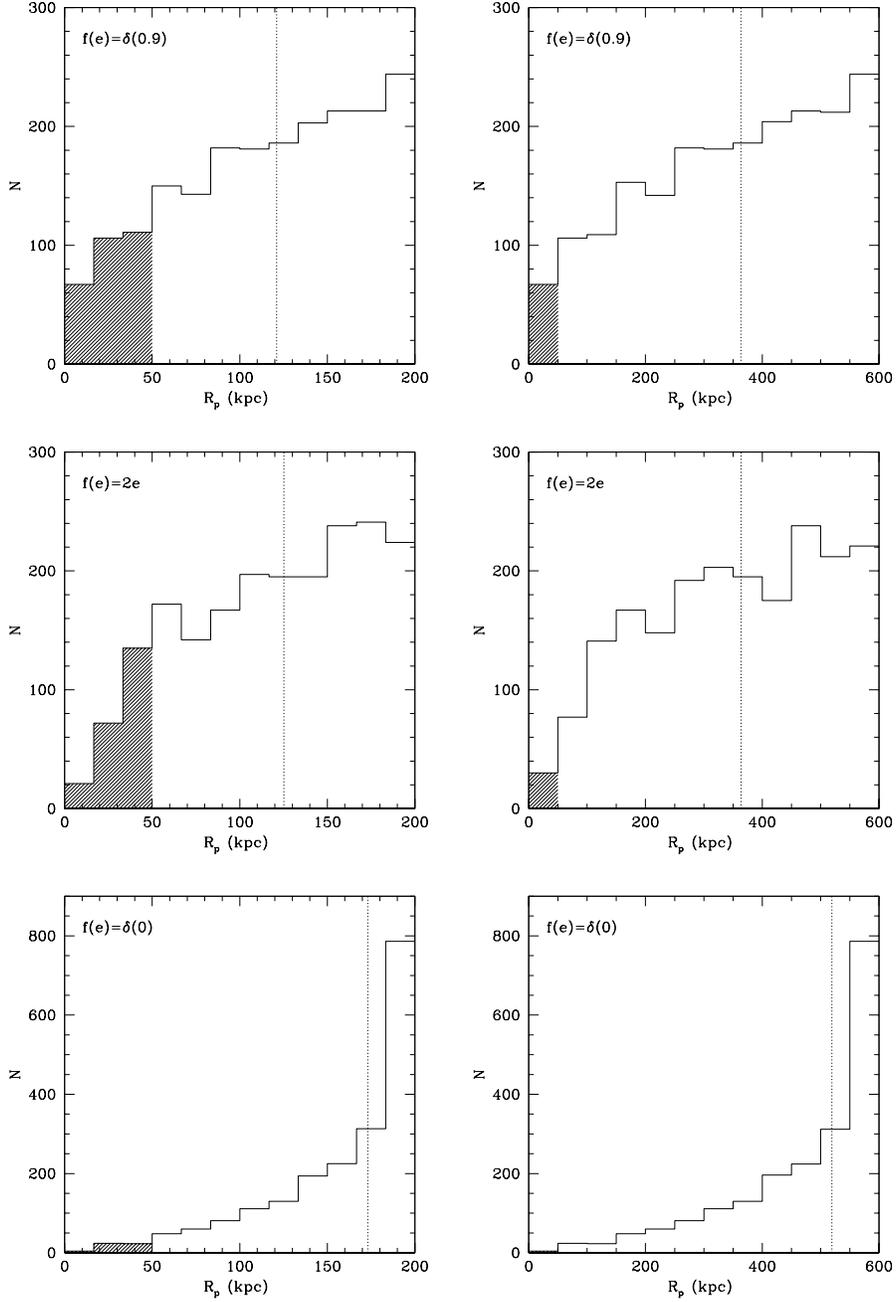}
%
%
      \caption{Distribution of projected separations $R_p$ for simulated pairs, 
with different eccentricity distributions $f(e)$ and apocentric separation 
$R_{apo}$. \emph{Left:} $R_{apo}=200$ kpc. 
\emph{Right:} $R_{apo}=600$ kpc. Dotted vertical line in all panels marks 
each sample's median $R_p$. Shaded areas show the close-pair population. 
              }
         \label{nrp}
   \end{figure}

Figure \ref{nrp} shows $N-R_p$ histograms for the two cases considered here, 
$R_{apo}=200$ and $600$ kpc, and for the three distributions of eccentricities, 
$f(e)=\delta (0.9), f(e)=2e$ and $f(e)=\delta(0)$. Shaded areas represent 
close pairs as defined above. Sample median values of 
$R_p$ are shown in the figure and in Table \ref{nclose}. As expected, the least 
median $R_p$ occurs for $f(e)=\delta (0.9)$ and for the triangular 
distribution $f(e)=2e$, being essentially the same for both distributions. 
It amounts to 60\% of $R_{apo}$, meaning about $120$ and $360$ kpc respectively. 
In the simulations of pairs in circular orbits, the median $R_p$ is exactly 87\% 
of $R_{apo}$. It can be noted from Fig. \ref{nrp} that the $R_p$ distributions 
are the same irrespective of $R_{apo}$ for all eccentricities. Such an overall 
behavior is expected due to self-similarity, since the dynamical problem is 
the same --- the sole 
difference on the orbit length scale. There is a slight fluctuation, however, in the 
triangular eccentricity distribution case (Fig. \ref{nrp}, \emph{middle panels}), 
which is due to the additional degree of freedom introduced by the random choice 
of the pair eccentricity. But even here $R_p$ distributions are the same, 
with a noticeable decrease of pairs at small projected separations as compared 
to the high-eccentricity simulations (Fig. \ref{nrp}, \emph{top panels}). 
%
%
%

The important difference between 
the distributions with $R_{apo}=200$ and $600$ kpc is the percentage 
of close pairs ($R_p\le50$ kpc). According to Table \ref{nclose}, 14\% of the 
total population of pairs with $f(e)=\delta(0.9)$ and $R_{apo}=200$ kpc are 
close pairs, a figure that decreases to 3\% if $R_{apo}=600$ kpc. This is a ready
consequence of the much wider phase space available for pairs with the larger 
$R_{apo}$.

As seen above, all simulations have large median $R_p$, even greater than 
$2\times R_{p,max}$, which shows that at least half of simulated pairs have 
spatial separations much greater than $R_{p,max}$. This is further confirmed 
by the data shown in Table \ref{rplus}, which represent the most important 
result, as far as the classification of close pairs is concerned:  
more than half of the simulated pairs with $R_p\le$ 50 kpc have 
three-dimensional separations greater than 50 kpc.

If real binaries are predominantly in circular orbits, 100\% of  pairs have
--- from input --- spatial separation of $R>$ 50 kpc, and close pairs are
thus very few, less than 3\% of the simulated population. The firm
conclusion from these results is that apparent separations are not
bona-fide indicators of spatial closeness. Close pairs with $R_{apo}<200$ kpc, 
not included in the simulations, would not substantially alter the conclusion.

In the next section, a classification scheme is put forward whose main purpose is 
to discriminate close pairs against their \emph{real} physical dynamical
character. The classification is a pre-requisite for any further
astrophysical investigation of the properties of galaxies in close pairs.

\section{The classification scheme}
The ground property in the classification scheme is the strength of tidal 
interaction between pair galaxies. Table \ref{clas} shows the four 
suggested classes of close pairs. The main goal is clearly discriminating
between pairs that are interacting through tidal forces and those whose
galaxies do not have their internal structure substantially affected by
the mutual differential gravitational interaction. 
The strength of tidal activity between two extended mass distributions, 
with centers of mass separated by $R$ is --- grossly speaking --- inversely 
proportional to $R^3$, thus justifying the consideration of tidal activity 
as the crucial property for a classification scheme. As seen in 
Table \ref{clas}, types CP I --- mergers --- and CP IV --- unphysical 
pairs --- are extreme types. Type I pairs are evolved type II pairs, and type 
IV pairs are misclassified close pairs, i.e., optical pairs.

Any study of the effects of closeness on the properties of galaxy pairs must be 
preceded by an observational selection procedure in order to classify the sample 
under study into the four classes. In the next section suggestions for the 
required observational programs are given. 

\section{Observational strategy for classifying close pairs}

\subsection{CP I and CP IV}
The CP I types are easily sorted by looking to their optical disturbed morphology. 
Individual galaxies are hardly seen as separate entities. The CP IV types, on  
the other hand, are selected by their discordant redshifts. A very 
conservative lower limit to the line-of-sight galaxy velocity difference 
in the pair of 300 km s$^{-1}$ may be 
adopted  as an initial guess to distinguish optical pairs. 
This limit corresponds to the relative velocity of two $10^{12} M_\odot$ 
(visible plus dark) galaxies on a circular orbit whose centers of mass are 
separated by 100 kpc. A detailed study of redshift asymmetries in the pair 
sample under investigation might be used in order to evaluate the contamination 
by optical companions. Valtonen \& Byrd \cite{valt} find that physical pairs 
exhibit an equal number of positive or negative redshifts relative to the 
primary while that is not true for non physical optical pairs.

The most difficult types to classify are CP II and CP III, which require a 
specially suitable observational strategy that takes into account the 
kinematical consequences of tidal forces.

\subsection{Tidal kinematics}
A straight application of the impulse approximation (Binney \&
Tremaine 1987) for tidal effects in high-speed 
encounters (``fly-bys'') of extended objects leads to
symmetrical perturbations. That is, two test particles, at symmetric
positions with respect to the perturbed object center, acquire velocity
increments of the same magnitude and opposite direction (Binney
\& Tremaine, eq. [7-54], p. 438). This will result in symmetric velocity
profiles. A perturbed disk galaxy would thus exhibit no signature of
tidal forces in its rotational profile.

Although the impulse approximation has been successfully shown to be
consistent with numerical experiments for mass and energy exchanges
between interacting spherical galaxies (Aguilar \& White 1985), it
does not hold here. Bound pairs are mid way between the fly-by
encounters in which the impulse approximation does apply and pairs that are
quickly evolving through a merger phase. Concerning the latter case,
Mihos et al. \cite{midh} made an \emph{N}-body merger model
of NGC 7252, a galaxy with conspicuous tidal tails. The initial
pre-merger galaxies have symmetrical rotation curves that evolve along
the merger process to a final asymmetrical velocity profile
(see their Fig. 6). Hence, one expects that at perigalacticon bound pairs
will show slightly --- but noticeably --- distorted
velocity profiles before they ultimately enter a merging phase. 

Furthermore, Barton et al. \cite{bart1} have shown, using \emph{N}-body 
simulations of galaxy encounters, that tidal interactions induce 
noticeable distortions on the rotation curves of spiral galaxy models 
(see their Fig. 1).

The asymmetry of the velocity profile of both galaxies at closest approach 
will be used here as an indication of {\it strong} --- loud --- tidal
forces, therefore implying spatial proximity. This is the rationale  underlying 
the discussion of the kinematics of CP II and CP III classes below. 

\subsection{CP II and CP III}
\begin{figure}
   \centering
\includegraphics[width=9cm]{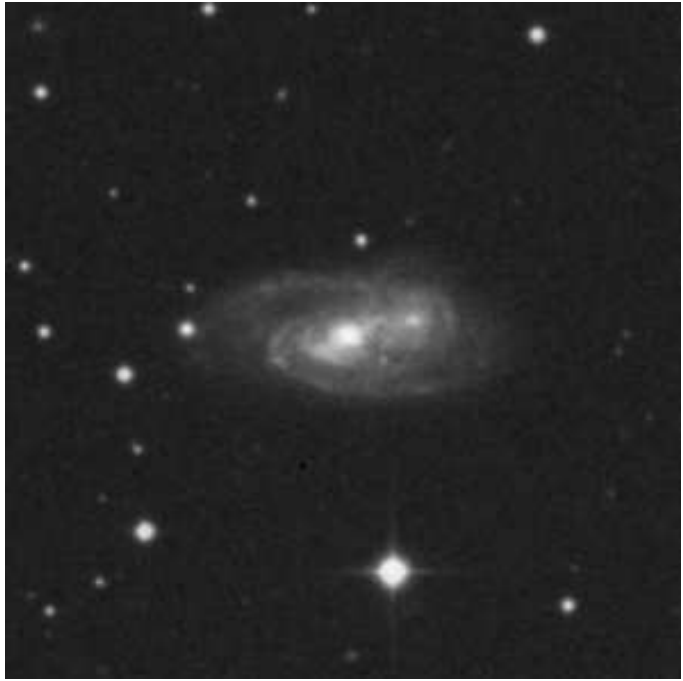}
%
%
      \caption{\emph{B}-band image from the POSS-II Digital Sky Survey of the close pair 
UGC 3995A+B. The smaller galaxy B is seen in front of the disk of the larger 
component. See also Marziani et al. \cite{marz}. North is up, and east is left.  }
         \label{cpg140}
   \end{figure}

The most confident criteria for CP II and CP III classification are: (1) broadband 
photometry, (2) kinematics from optical single-slit spectroscopy, and (3) global 
H {\sc i} spectral line low-resolution velocity profiles. These are intentionally 
designed to be low cost procedures for the classification. Strong tidal 
forces affect the outcome of all three sorts of observations. Ideally, 
all three should be used, as, for example, in Marziani et al. \cite{marz}. They 
studied the close pair UGC 3995A+B (CPG 140, in Karachentsev 1972, 1987), whose 
POSS-II \emph{B}-band image is shown in Fig. \ref{cpg140}.

Galaxy centers are separated by approximately 30\arcsec, and
the smaller component B is seen in front of the spiral disk of component A. 
Marziani et al. \cite{marz} proceeded to broadband photometry decomposition and found 
unperturbed individual solutions for both galaxies. Optical spectroscopy 
reveals symmetrical kinematics relative to the galaxy centers. That is, the 
rotation-curve amplitudes are the same on the receding and approaching sides 
of both galaxies. The H {\sc i} global profile is symmetrical as well. The optical 
velocity difference is 20 km s$^{-1}$ (internal error of 18 km s$^{-1}$), and the 
H {\sc i} velocity difference is 1 km s$^{-1}$, from Sulentic \& Arp \cite{sa83}. These 
observations suggest that the pair is a CP III, a ``tide-quiet'' close pair. 
seen at or close to apocentric orbital phase, near turnover. 
It is worthwhile mentioning that the orbital 
interpretation above diverges from Marziani et al. They claim that the pair 
is seen either before or after their closest approach. This seems unlikely, however 
because in this case tidal perturbations would be evident in the
observations. The classification scheme relies on the 
assumption that tidal activity impresses asymmetrical signatures on morphology and 
kinematics. 

\begin{figure}
   \centering
\includegraphics[width=9cm]{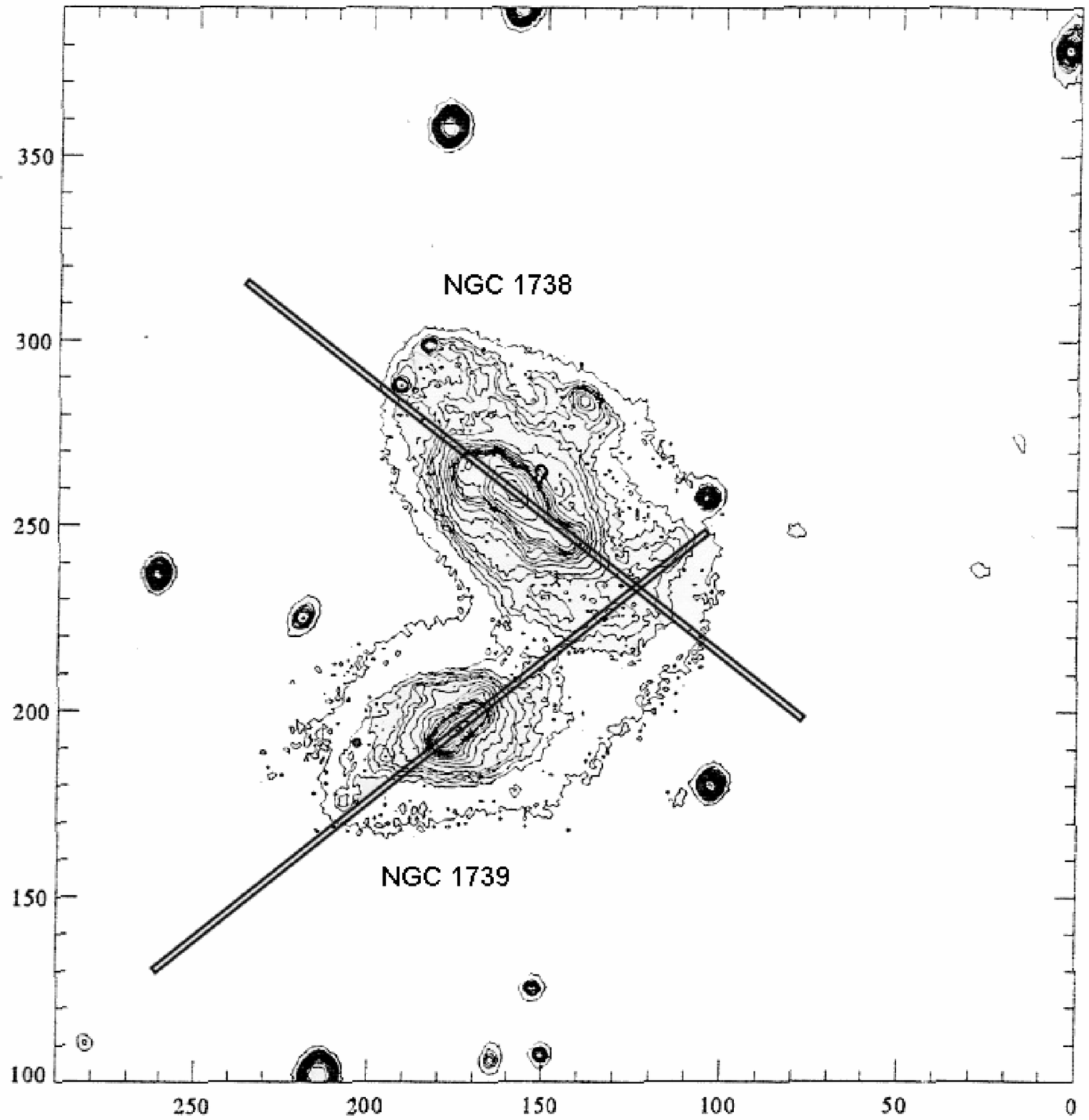}
%
%
      \caption{\emph{R}-band isophote map of SBG 249 from Carvalho \& Soares (2007, in 
preparation). NGC 1738 (ESO-LV 552490) is the background galaxy. Slits are 
128\arcsec\ long. North is up, and east is left. }
         \label{sbg249map}
   \end{figure}

\begin{figure}
   \centering
\includegraphics[width=12cm]{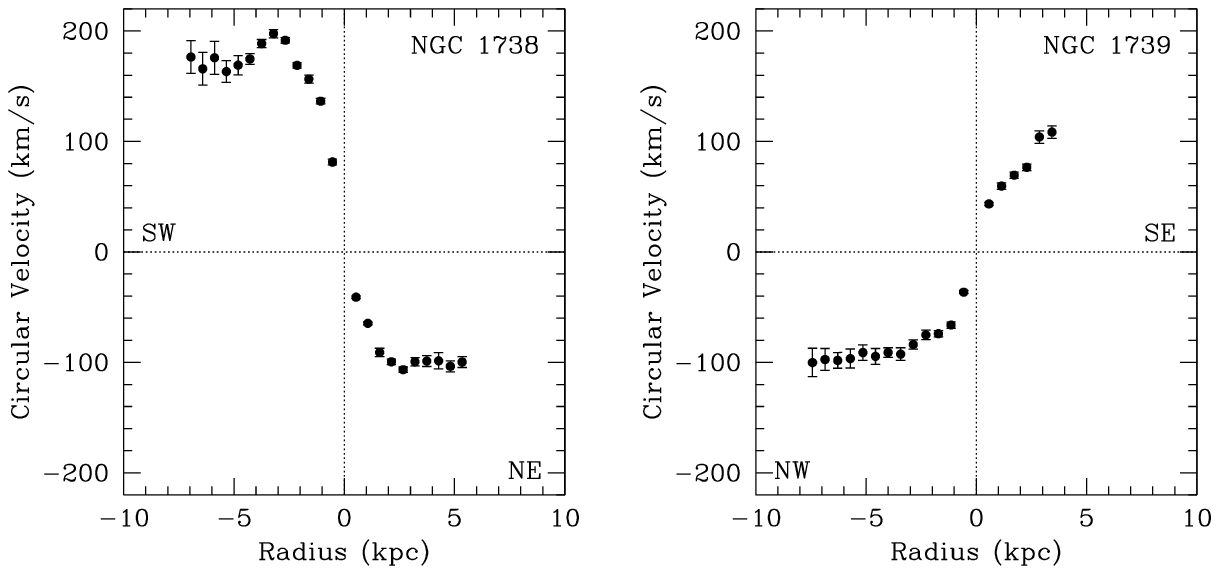}
%
%
      \caption{Rotation curve of NGC 1738 (\emph{left}) and NGC 1739 
       (Carvalho \& Soares 2007, in preparation).  The southwestern side (SW) of 
       the NGC 1738 disk and the northwestern side (NW) of NGC 1739 are partly 
       overlapped. }
         \label{sbg249}
   \end{figure}

A counterexample to UGC 3995, namely, a ``tide-loud'' close pair, is SBG 249 
(Soares et al. 1995). Long-slit spectroscopy has been performed and analyzed 
by Carvalho \& Soares (2007, in preparation). The galaxies are NGC 1738 
(ESO-LV 5520490) and NGC 1739 (ESO-LV 5520500). They have partially 
overlapped disks. Two spectra were obtained with the Double Spectrograph 
instrument mounted at the Palomar 5 m telescope, on 1998 February 27--28. 
Figure \ref{sbg249map} shows an isophote map 
of the pair, with the 128\arcsec\ long slits superposed. The derived rotation 
curves are shown in Fig. \ref{sbg249} (both figures from Carvalho \& Soares 
2007, in preparation). 
%

The asymmetry is apparent in NGC 1738 and hinted at in NGC 1739. The southwestern 
side of NGC 1738 rotates almost twice as much as the northeastern side; that is, 
its closest side to NGC 1739 rotates at a greater speed than the farthest side.
The northwestern side of NGC 1739 flattens out at about 100 km s$^{-1}$, while the
southeastern side seems to be rising up at the last measured point. It is
clear that the rotation profiles are strongly affected by their mutual
tidal influence. On the other hand, the spectrum of the overlapped portion 
of the system is clearly disentangled showing that both disks rotate quite 
independently and, as can be seen in Fig. \ref{sbg249}, in opposite 
directions. This also rules out classification as a CP I, in spite of
the disturbed morphological appearance of the pair.

 The line of sight velocity difference is 60 km s$^{-1}$ (internal
error of 20 km s$^{-1}$). Since they are low-mass galaxies, this is consistent
with SBG 249 being a CP II seen at or close to perigalacticon in an
eccentric orbit (cf. the Appendix in Soares 1996). Additional
broadband photometry and global H {\sc i} observation are planned for the pair. 

The strength of the asymmetry in the velocity profile might depend on the disk 
rotation sense relative to the orbit sense. Howard et al. \cite{how} performed 
detailed simulations of tidally induced structures in disk galaxies. They 
discuss morphological disturbances and find that retrograde encounters 
are less effective in producing disturbances. If the result applies also to 
kinematical disturbances then pairs that pass close enough to be CP II --- 
tide-loud pairs --- might  resemble  CP III types if the galaxy's rotation sense 
is opposite to the orbit  sense. These cases require careful analysis of 
\emph{both} broadband photometry and pair kinematics as was the case for CPG 140 
discussed above.

\section{Conclusion}
Monte Carlo simulations of close pairs of galaxies show that a projected
separation restrictive criterion does not guarantee closeness in space.
Accordingly, a simple classification scheme of close pairs is proposed
and an observational strategy is suggested for sorting out a sample
before any kind of astrophysical investigation of their properties is made.
The classification is based on kinematical and dynamical criteria and on the 
pair's overall morphological appearance. The required observations are thus 
broadband photometry and optical and H {\sc i} 21-cm line spectroscopy. The 
latter is of course only suitable for pairs with late-type galaxies.

For a sample of close pairs, defined initially by their closeness
in the plane of sky, the three kinds of observations are 
sufficient in ascribing to each pair one of the classes suggested. Classes
are defined by the strength of present tidal activity and optical morphology.
They are mergers (CP I), tide-loud pairs (CP II), tide-quiet pairs (CP III), and
physically unbound optical pairs (CP IV).

\begin{acknowledgements}
I thank David Balparda de Carvalho for helpful discussions on the close-pair 
subject and for a careful reading of the manuscript. The anonymous referee is 
gratefully acknowledged for useful comments and suggestions to improve the 
text. The Second Palomar Observatory Sky Survey was made by the 
California Institute of Technology with funds from the National Science Foundation, 
the National Geographic Society, the Sloan Foundation, the Samuel Oschin 
Foundation, and the Eastman Kodak Corporation. I thank the Brazilian agency 
FAPEMIG for partial support. 
\end{acknowledgements}

\end{document}